\documentclass[12pt]{article}
\global\arraycolsep=2pt 
\input{epsf}
\begin{document}
\makeatletter
\def\fmslash{\@ifnextchar[{\fmsl@sh}{\fmsl@sh[0mu]}}
\def\fmsl@sh[#1]#2{%
  \mathchoice
    {\@fmsl@sh\displaystyle{#1}{#2}}%
    {\@fmsl@sh\textstyle{#1}{#2}}%
    {\@fmsl@sh\scriptstyle{#1}{#2}}%
    {\@fmsl@sh\scriptscriptstyle{#1}{#2}}}
\def\@fmsl@sh#1#2#3{\m@th\ooalign{$\hfil#1\mkern#2/\hfil$\crcr$#1#3$}}
\makeatother
\thispagestyle{empty}
\begin{titlepage}

\begin{flushright}
hep-th/0008189 \\
LMU 13/00\\
\today
\end{flushright}

\vspace{0.3cm}
\boldmath
\begin{center}
\Large\bf Hidden Supersymmetry
\end{center}
\unboldmath
\vspace{0.8cm}

\begin{center}
  {\large X. Calmet} \\
  {\sl Ludwig-Maximilians-University, Sektion Physik} \\
  {\sl Theresienstr.37, D-80333 Munich, Germany.}
\end{center}

\vspace{\fill}

\begin{abstract}
\noindent
Inspired by the concept of complementarity, we present a illustrative
model for the weak interactions with unbroken gauge symmetry and
unbroken supersymmetry.  The observable particles are bound states of
some more fundamental particles. Supersymmetry is broken at the
macroscopic scale of the observable particles by a discrete symmetry
but remains exact at the scale of the fundamental particle and is thus
hidden.  This provides a link between theories at very high energies
and the observed particle physics. Supersymmetric particles are
confined in usual matter.
\end{abstract}
to appear in Physics Letters B 
\end{titlepage}

\section{Introduction}
In seminal papers \cite{tHooft1,tHooft2}, 't Hooft has shown that in a
gauge theory with a Higgs boson in the fundamental representation of
the gauge group there is no fundamental difference between the theory
in the Higgs phase, i.e, with a gauge symmetry broken by means of the
Higgs mechanism and the theory in the confinement phase i.e, a theory
with confined gauge charges. This property is known as the
complementarity principle \cite{Osterwalder:1978pc}. The fundamental
difference between the electroweak theory and QCD is that in the
electroweak sector one has a large parameter which allows perturbative
calculations.

Recently we have used this complementarity to build an alternative to
the standard model using a $SU(2)_L \otimes U(1)_Y$ gauge group with
$SU(2)_L$ confinement \cite{CF}.  We have clarified how to incorporate
quantum electrodynamics in that kind of models and studied the
physical consequences of the assumption that the electroweak
interactions might be described by the confinement phase.  In that
case all phenomena in particle physics are described by exact gauge
theories. If nature is such that its fundamental Lagrangian has the
maximal number of allowed symmetries, it is natural to assume that
supersymmetry could also be an exact symmetry of this Lagrangian.
Supersymmetry is a crucial aspect in particle physics, it is a
desirable feature of many high energy theories like some variants of
grand unified theories. It is the missing link between some theories
at very high energies and low energy particle physics.

It is thus meaningful to design mechanisms that explain why
supersymmetry is unobserved. A possibility is that supersymmetry is
broken. This leads to models such as the minimal supersymmetric
standard model (MSSM). We propose an alternative point of view. If the
electroweak interactions are described by a confining theory, the
microscopic theory can be supersymmetric but this symmetry is then
hidden at the macroscopic scale of fermions and electroweak bosons.
In other words we will break supersymmetry at the macroscopic scale
without breaking it at the scale of fundamental particles thus
providing a link between some theories at very high and low energy
particle physics.

In composite models, supersymmetry is not necessary to solve the
hierarchy problem because the Higgs boson is not a fundamental
particle but it remains important to have a supersymmetric theory to
reach the unification of the coupling constants at the unification
scale.

We then consider a supersymmetric extension of the model for the
electroweak interactions proposed in \cite{CF} with broken
supersymmetry at the fundamental level.

\section{A supersymmetric toy model}

We shall consider a illustrative model with the gauge group $SU(2)_L$ and
unbroken $\!N\!=\!1\!$ supersymmetry.  The situation in a gauge theory
with unbroken supersymmetry is very similar to that of the confinement
phase in a non-supersymmetric theory. We assume that there is a
$SU(2)_L$ confinement: all physical particles are $SU(2)_L$ singlets.
We have the following particle spectrum: the right-handed fermions
$e_R$, $u_R$, $d_R$ and their superpartners $\tilde{e}_R$,
$\tilde{u}_R$, $\tilde{d}_R$. The right-handed particles are the usual
right-handed leptons and quarks of the standard model and their
superpartners, whereas the left-handed doublets are bound states of
some more elementary particles. The fundamental $SU(2)_L$ fields
(D-quarks) are:
\\
\begin{tabular}{lll}
leptonic D-quarks & $l_i=  \left(\begin{array}{c} l_1 \\ l_2 \end{array}
\right )$  &  (fermions)  \\
& & \\
 hadronic  D-quarks   & $q_i= \left(\begin{array}{c}q_1 \\ q_2\end{array}
\right )$   &  (fermions, $SU(3)_c$ triplets) \\ 
& & \\
scalar D-quarks  &  $h_i= \left(
  \begin{array}{c}
  h_1 \\ h_2
  \end{array}
\right )$ & (bosons).
\end{tabular}
\\
Notice that in order to cancel the anomalies we would have to
introduce a second scalar doublet. We discard this problem as our aim
is only to present a toy model to emphasize our idea. We then have the
superpartners
\\
\begin{tabular}{lll}
leptonic D-squarks & $\tilde{l}_i=  \left(\begin{array}{c}
    \tilde{l}_1 \\ \tilde{l}_2 \end{array}
\right )$  &  (bosons)  \\
& & \\
hadronic  D-squarks   & $\tilde{q}_i= \left(\begin{array}{c}
    \tilde{q}_1 \\ \tilde{q}_2\end{array}
\right )$   &  (bosons, $SU(3)_c$ triplets) \\ 
& & \\
scalar D-squarks  &  $\tilde{h}_i= \left(
  \begin{array}{c}
  \tilde{h}_1 \\ \tilde{h}_2
  \end{array}
\right )$ & (fermions).
\end{tabular}
\\
We shall refer to the theory involving the D-quarks and the
D-squarks as the microscopic theory. At the macroscopic level i.e,
the theory of bound states, a large number of $SU(2)_L$ invariant bound
states can be identified.  We see that bound states of different
particles can have the same quantum numbers. For example, the neutrino
can be identified with the bound state $\bar h l$ but also with the
bound state $\bar{\tilde{h}} \tilde{l}$.  It will thus be a
superposition of both bound states. This can be
applied to the rest of the known particles. The left-handed
fermions, normalized in the appropriate way, are defined as
follows. We have the leptons

\begin{eqnarray} \label{eq1}
\mbox{left-handed neutrino} \ \nu_L &=& \frac{1}{F}
\left ( (\bar h l) +  (\bar{\tilde{h}} \tilde{l}) \right ) \\
\mbox{left-handed electron} \ e_L &=& \frac{1}{F}
\left ( (\epsilon^{ij} h_i l_j)
+  (\epsilon^{ij} \tilde{h}_i \tilde{l}_j) \right ) \nonumber
\end{eqnarray}
where $F$ is a numerical, to be specified, normalization factor. The
quarks are also bound states

\begin{eqnarray} \label{eq2}
\mbox{left-handed up quark} \ u_L &=& \frac{1}{F}
\left ( (\bar h q) +  (\bar{\tilde{h}} \tilde{q}) \right ) \\
\mbox{left-handed down quark} \ d_L &=& \frac{1}{F}
\left ( (\epsilon^{ij} h_i q_j)
+  (\epsilon^{ij} \tilde{h}_i \tilde{q}_j) \right ). \nonumber
\end{eqnarray}

The Higgs and electroweak bosons are bound states of scalar D-quarks and
their superpartners:
\begin{eqnarray} \label{eq3}
\mbox{Higgs field} \ \phi &=& \frac{1}{2 F}
\left ( (\bar h h) + \beta  (\bar{\tilde{h}} \tilde{h}) \right) \\ \nonumber
\mbox{electroweak boson} \ W^3_\mu&=& \frac{2 i}{g F^2}
\left ( (\bar h D_\mu h) +  \beta (\bar{\tilde{h}} D_\mu \tilde{h}) \right ) \\
\mbox{electroweak boson} \ W^-_\mu&=& \frac{\sqrt{2} i}{g F^2}
\left ( (\epsilon^{ij} h_i D_\mu h_j) + \beta
(\epsilon^{ij} \tilde{h}_i D_\mu \tilde{h}_j) \right ), \nonumber
\end{eqnarray}
where $D_\mu$ is the covariant derivative of the gauge group $SU(2)_L$
involving the gauge bosons $B^a_\mu$ and $g$ is the gauge coupling of
this group. The second charged $W$ boson $W^+$ is defined as
$(W^-)^\dagger$. A simple dimensional analysis shows that a constant
$\beta$ with dimension $-1$ has to appear. This constant is a priori
unknown but the only scale of the theory being $F$, we could impose
$\beta=1/F$.  This apparently arbitrary choice is not a drawback for
the theory as we will see that only the terms containing a scalar
D-quark doublet will be relevant. 

The problem is to know whether a particle and its superparticle will
belong to the same supermultiplet, i.e, if they have the same mass. It
is a difficult question as dynamical effects can contribute to the
masses.  For example, the masses of the electroweak bosons are to a
large extent dominated by dynamical effects. Once we have introduced a
second Higgs doublet, we have the same gauge group and the same
particle contain as in the MSSM, dynamical supersymmetry breaking is
thus possible.  There are two possibilities: either the masses of, for
example, an electroweak boson and of the corresponding superparticle
are identical and supersymmetry is unbroken at the macroscopic level
or they are different because of dynamical effects and supersymmetry
is dynamically broken. This possibility can't be excluded, but in the
sequel we assume that these particles indeed form a supermultiplet.
Thus, an electron is the superpartner of a selectron. Lattice
simulations could test the dynamical behavior of such a model.

All the particles we have identified up to this point are those
appearing in the standard model. We can also identify the
bounds states corresponding to the macroscopic superparticles. For
example, we have
\begin{eqnarray}
\mbox{selectron} \ 
\tilde e &=& \frac{1}{F}
\left ( (\epsilon^{ij} h_i \tilde{l}_j)
+ \beta (\epsilon^{ij} \tilde{h}_i l_j) \right ) \nonumber
\end{eqnarray}
for the left-handed selectron.

The complementarity principle was established in the framework of a
non-supersymmetric theory with a single Higgs boson doublet. This
principle requires that the coupling constants between the bounds
states and the electroweak bosons are the same in the Higgs phase and
in the confinement phase. In the case of a non-supersymmetric model
\cite{CF}, 't Hooft proposed that the confinement is due to vortices
\cite{tHooft1}. This means that we have a confinement with a weak
coupling constant which avoids the problems due to chiral symmetry
breaking \cite{Hsu:1993zc}.

In a supersymmetric model the situation is more complex since the
theory is richer. Nevertheless the situation in such a theory is very
similar to that of the confinement phase in a non-supersymmetric gauge
theory.  The question is whether our microscopic model which is
supersymmetric will have a supersymmetric macroscopic spectrum.  A
lattice study of the vacuum structure and of the dynamical behavior of
our model would be useful to answer this question. As long as this as
not been done some place is left for speculation.

A discrete symmetry could explain why nature selects, at least at low
energy, only the particles.  We introduce a mechanism similar to the
so-called R-parity. We assign a new quantum number to the particles.
We call this new quantum number S-parity. The D-quarks are assigned
S-parity +1, whereas the D-squarks are assigned S-parity -1. We then
assume that the bound states appearing in nature have S-parity +1.

This selection rule shifts the masses of the superparticles to very
high energies. In other words we break supersymmetry at the
macroscopic level by imposing a discrete symmetry but it remains
intact at the microscopic level. It is thus clear that superparticles
corresponding to the left-handed particles, to the Higgs sector and to
the electroweak bosons will not be observable at least at low energy.
In that case, we expect that a confining theory describes the weak
interactions correctly.  Imposing this selection rule which is
motivated by the apparent absence of superparticles in nature at low
energy is not trivial as it would be in the case of the MSSM because
the fundamental D-squarks are confined in usual matter. It would not
be very surprising if this S-parity was broken in nature, as there are
already many examples of broken discrete symmetries. But, at this
stage it remains a speculation, which could be tested on the lattice.

That scenario is useful in the case of a grand unified theory.  If
there is a deconfinement phase at the scale of a few TeV,
supersymmetry is realized above that scale and the coupling constants
unification takes place at the unification scale, but supersymmetry
remains hidden at low energy under this deconfinement phase.  Two
scenarios are conceivable, the mass scale of the superparticles is
below the deconfinement scale, in which case one will observe
superparticles but the theory is not explicitly supersymmetric until
one reaches the deconfinement scale.  Another possibility is that the
mass scale for the superparticles is above the deconfinement scale in
which case the particle spectrum would suddenly become supersymmetric
above the deconfinement scale. This feature allows to test our idea.

Even if supersymmetry was broken by dynamical effects, it might still
be necessary, if the mass splitting was not sufficiently large, to
introduce this S-parity for phenomenological reasons.

\section{Back to known particles}

It remains to show that the definitions for the fields indeed describe
the observed particles. We use the unitary gauge for the scalar doublet
\begin{eqnarray}
   h_i=\left ( \begin{array}{c} F+h_{(1)} \\ 0 \end{array}\right).
\end{eqnarray}

The parameter $F$ is a real number.  If $F$ is sufficiently
large we can perform a $1/F$ expansion for the fields defined
previously, we then have

\begin{eqnarray} 
 \nu_L&=&l_1+\frac{1}{F} \left ( h_{(1)} l_1
   + \bar{\tilde{h}} \tilde{l} \right )\approx l_1  \\
  e_L&=&
  l_2+\frac{1}{F} \left ( h_{(1)} l_2 +
\epsilon^{ij} \tilde{h}_i \tilde{l}_j \right )
    \approx l_2 
  \nonumber \\
   u_L&=&q_1+\frac{1}{F} \left ( h_{(1)} q_1
 + \bar{\tilde{h}} \tilde{q} \right )
   \approx q_1 
   \nonumber \\
     d_L&=&q_2 
     +\frac{1}{F} \left ( h_{(1)} q_2
+
\epsilon^{ij} \tilde{h}_i \tilde{q}_j \right )
       \approx q_2
     \nonumber \\
     \phi&=&h_{(1)}+\frac{F}{2} +\frac{1}{2 F}  \left (
     h_{(1)} h_{(1)} + \beta
\bar{\tilde{h}} \tilde{h} \right)
     \approx h_{(1)}+\frac{F}{2}
     \nonumber \\ \nonumber
     W^3_{\mu}&=&\left ( 1 + 
     \frac{h_{(1)}}{F} \right)^2 B^3_{\mu} + \frac{2 i}{g F} \left (1+ 
     \frac{h_{(1)}}{F} \right) \partial_{\mu} h_{(1)}  \\ && \nonumber  
+\frac{2 i\beta}{g F^2} \left ( \bar{\tilde{h}} D_\mu \tilde{h} \right )
   \approx B^3_{\mu}
    \nonumber \\ \nonumber
   W^-_{\mu}&=& \left ( 1 +
     \frac{h_{(1)}}{F} \right)^2 B^-_{\mu} +
\frac{\sqrt{2} i\beta}{g F^2}
\left ( \epsilon^{ij} \tilde{h}_i D_\mu \tilde{h}_j \right )
   \approx B^-_{\mu}. \
\end{eqnarray}
As also done by 't Hooft \cite{tHooft1,tHooft2}, we assume that the
only particles which are stable enough to be observable at presently
accessible energies are those containing the scalar doublet $h$, those
are the only fields who survive to the expansion, and we consider the
terms suppressed by a factor $1/F$ as being irrelevant.  Therefore the
spectrum of this theory is, for the left-handed sector, identical to
the spectrum of the standard model. Nevertheless we are not able to
hide the superpartners of the right-handed particles at this stage.
Supersymmetry is apparently broken in the left-handed sector but in
fact it remains unbroken at the microscopic level of the theory.

\section{The MSSM}

In this section, we assume that the complementarity principle remains
valid for supersymmetric theories once soft breaking terms have been
introduced. The model in the confinement phase corresponding to the
minimal supersymmetric standard model can easily be obtained by
requiring that supersymmetry is broken by usual means at the level of
the fundamental D-quarks and D-squarks. A second Higgs doublet $k$ and
the corresponding superparticle $\tilde k$ can be introduced without
any difficulty, and we basically have to replace $h$ and $\tilde h$ by
$s=h + i \sigma_2 k^*$ and $\tilde s= \tilde h + i \sigma_2 \tilde
k^*$ in the definitions of the fermions, superfermions, electroweak
bosons and of their superpartners. The gauge is fixed in such a way
that $s$ takes the form $s=(F+ h_{(1)} + k_{(1)},0)$, where
$F=F_1+F_2$, $F_1$ corresponding to the scalar doublet $h$ and $F_2$
to the scalar doublet $k$. We then have
\begin{eqnarray}
h= \left(\begin{array}{c}
    F_1 + h_{(1)} + i \chi^0 \\  - \phi^- \end{array}
\right ), \ \
k= \left(\begin{array}{c}
    - \phi^+ \\ F_2 + k_{(1)} + i \chi^0 \end{array}
\right ).
\end{eqnarray}
 
We can define the five Higgs bosons
\begin{eqnarray}
\mbox{$CP$ even Higgs boson} \
\phi_1&=& \frac{1}{2 F_1} \left ({\bar h h}
\right)
  = h_{(1)}+ \frac{F_1}{2} + {\cal O}\left(\frac{1}{ 2 F_1}\right)
\\ \nonumber
\mbox{$CP$ even Higgs boson} \
\phi_2&=& \frac{1}{2 F_2} \left ({\bar k k}
\right)
  = k_{(1)}+ \frac{F_2}{2} + {\cal O}\left(\frac{1}{2 F_2}\right)
\\\
\mbox{$CP$ odd Higgs boson} \
i \chi&=& \left (
\frac{1}{2 F} (\bar s h +  \epsilon^{ij} s_i k_j) 
-\frac{1}{2 F_1} (\bar h h)- \frac{1}{2 F_2} (\bar k k) \right)
\nonumber \\ &=& i \chi + {\cal O}\left(\frac{1}{ 2 F_1}\right)
+ {\cal O}\left(\frac{1}{2 F_2}\right)
\nonumber 
\\
\mbox{charged Higgs boson} \
\phi^+&=& \frac{-1}{F} \left ({\bar s k}
\right)
  = \phi^+ + {\cal O}\left(\frac{1}{ F}\right)
\nonumber
\\
\mbox{charged Higgs boson} \
\phi^-&=& \frac{-1}{F} \left ({\epsilon^{ij} s_i h_j}
\right)
  = \phi^- + {\cal O}\left(\frac{1}{ F}\right).
\nonumber 
\end{eqnarray}

The superpartners of these Higgs bosons can be obtained in a similar
way.  The model presented in \cite{CF} is thus compatible with a
supersymmetric extension provided that both $F_1$ and $F_2$ can be
chosen to be large. This model has the same vertices as the MSSM and
the same particle contain. As in the case of the non-supersymmetric
model, we expect that radial and orbital excited versions of the known
particles will appear.

\section{Conclusions}

We have considered a toy model with $SU(2)_L$ confinement and hidden
supersymmetry in the left-handed sector. Supersymmetry is broken at
the macroscopic level by a discrete symmetry. The first step towards a
realistic model is to include a second Higgs doublet. It can be done
without major difficulties as has been shown in the last section.

This model can be extended to a model with a $SU(3)_c \otimes SU(2)_R
\otimes SU(2)_L \otimes U(1)_Y$ gauge group with two Higgs doublets
for each $SU(2)$ sector.  Once this extension has been done, we can
hide supersymmetry completely at the microscopic level for the
$SU(2)_R \otimes SU(2)_L$ sector, assuming a $SU(2)_R \otimes SU(2)_L$
confinement.  Supersymmetry would have to be broken by usual means for
the two remaining gauge groups. The spectrum of the macroscopic theory
at low energy is then that of the standard model with ten Higgs
fields, i.e. five for each $SU(2)$ sector, 8 gluinos and a photino.

This model provides the missing link between low energy particle
physics and very high energy theories like grand unified theories.
Usual models with supersymmetry breaking are not able to explain a
small cosmological constant \cite{Witten:2000zk}. In our approach,
supersymmetry is not broken in the $SU(2)_L$ sector at the microscopic
level.  Thus the contribution of the energy of the fundamental vacuum
of that sector to the cosmological constant is vanishing. Our
mechanism could therefore help to explain a small or vanishing
cosmological constant.

Note that this model would nicely fit into a supersymmetric $SO(10)$
grand unified theory, which thus could be the fundamental theory of
D-quarks and D-squarks. It turns out that such a theory would be very
similar to the standard model if there is a confinement in the weak
interactions sector.

Finally, we described a supersymmetric extension of the model proposed
in Ref. \cite{CF} for the electroweak interactions with $SU(2)$
confinement. We have shown that this model is compatible with a
supersymmetric extension provided that the complementarity principle
remains valid for supersymmetric theories once soft breaking terms
have been introduced.

A detailed study of the vacuum structure of supersymmetric $SU(2)$
gauge theories and of their dynamical behavior by lattice simulations
would allow to determine whether these scenarios can be realized in
nature.

\section*{Acknowledgements}
The author would like to thank H. Fritzsch for discussions and for
reading this manuscript. He would also like to thank R. Dick, F.
Klinkhamer, A. Leike, J.  Pati, I. Sachs, E. Seiler and Z. Xing for
stimulating discussions.

\end{document}